\documentclass[12pt,preprint]{aastex}
\usepackage{graphicx}
\usepackage{amssymb}
\usepackage{natbib}
\usepackage{amsmath}

\newcommand{\myemail}{mat@igam.uni-graz.at}

\newcommand{\degree}{^{\circ}}

\begin{document}

\title{Characteristics of kinematics of a coronal mass ejection during the 2010 August 1 CME-CME interaction event}

\author{Manuela Temmer$^{1}$, Bojan Vr\v{s}nak$^{2}$, Tanja Rollett$^{1}$, Bianca Bein$^{1}$, Curt~A. de Koning$^{3}$, Ying Liu$^{4,5}$, Eckhard Bosman$^{6}$, Jackie~A. Davies$^{7}$, Christian M\"ostl$^{1,4,6}$, Tomislav \v{Z}ic$^{2}$, Astrid~M. Veronig$^{1}$, Volker Bothmer$^{7}$, Richard Harrison$^{8}$, Nariaki Nitta$^{9}$, Mario Bisi$^{10,11}$, Olga Flor$^{1}$, Jonathan Eastwood$^{12}$, Dusan Odstrcil$^{13}$, Robert Forsyth$^{12}$ }

\affil{Kanzelh\"ohe Observatory-IGAM, Institute of Physics, University of Graz,
Universit\"atsplatz 5, A-8010 Graz, Austria. email: \myemail}

\affil{Hvar Observatory, Faculty of Geodesy, University of Zagreb,
Ka\v{c}i\'{c}eva 26, HR-10000 Zagreb, Croatia}

\affil{NOAA Space Weather Prediction Center, Boulder Colorado, CO 80305, USA}

\affil{Space Sciences Laboratory, University of California, Berkeley, CA 94720, USA}

\affil{State Key Laboratory of Space Weather, National Space Science Center,
Chinese Academy of Sciences, Beijing, China}

\affil{Space Research Institute, Austrian Academy of Sciences, A-8042 Graz, Austria}

\affil{Institut f\"ur Astrophysik, G\"ottingen University, Friedrich-Hund Platz 1, D-37077 G\"ottingen, Germany}

\affil{RAL Space, Rutherford Appleton Laboratory, Harwell Oxford, Didcot OX11 0QX, UK}

\affil{Solar and Astrophysics Laboratory, Lockheed Martin Advanced Technology Centre, Palo Alto, CA 94304-1191, USA}

\affil{Institute of Mathematics and Physics, Aberystwyth University, Ceredigion SY23 3BZ, Wales, UK}

\affil{Center for Astrophysics and Space Sciences, University of California, San Diego, La Jolla, CA 92093-0424, USA}

\affil{The Blackett Laboratory, Imperial College London, London SW7 2AZ, UK}

\and

\affil{George Mason University/NASA Goddard Space Flight Center, Mail code 674, Greenbelt, MD 20771, USA}

\begin{abstract}
We study the interaction of two successive coronal mass ejections (CMEs) during the 2010 August 1 events using STEREO/SECCHI COR and HI data. We obtain the direction of motion for both CMEs by applying several independent reconstruction methods and find that the CMEs head in similar directions. This provides evidence that a full interaction takes place between the two CMEs that can be observed in the HI1 field-of-view. The full de-projected kinematics of the faster CME from Sun to Earth is derived by combining remote observations with in situ measurements of the CME at 1~AU. The speed profile of the faster CME (CME2; $\sim$1200~km~s$^{-1}$) shows a strong deceleration over the distance range at which it reaches the slower, preceding CME (CME1; $\sim$700~km~s$^{-1}$). By applying a drag-based model we are able to reproduce the kinematical profile of CME2 suggesting that CME1 represents a magnetohydrodynamic obstacle for CME2 and that, after the interaction, the merged entity propagates as a single structure in an ambient flow of speed and density typical for quiet solar wind conditions. Observational facts show that magnetic forces may contribute to the enhanced deceleration of CME2. We speculate that the increase in magnetic tension and pressure, when CME2 bends and compresses the magnetic field lines of CME1, increases the efficiency of drag.
\end{abstract}

\section{Introduction}

On 2010 August 1 a total of five coronal mass ejections (CMEs) could be identified. All CMEs were launched from source regions located on the Earth-directed hemisphere of the Sun, which made them potential geoeffective space weather events \citep[for an overview of all these CME events see][]{harrison12}. \cite{schrijver11} presented a detailed study focusing on the 2010 August 1 CME-associated solar surface activity using observations from the Solar Terrestrial Relations Observatory (STEREO) and the Solar Dynamics Observatory (SDO). They showed how complex and widely distributed the magnetic field line connections were over the solar surface that finally caused the initiation and evolution of the observed eruptions. In fact, the eruptions affected half of the solar corona which makes it necessary to analyze many aspects of these events in detail.

Two of the CME events on 2010 August 1 \citep[named as M- and L-tracks in the overview paper by][]{harrison12} were launched in quick succession from source regions located close together \citep{harrison12}. Furthermore, as we shall show, the first event was slower than the second event, hence, the CMEs are expected to interact with each other at a certain distance from the Sun. It has been suggested by \cite{gopalswamy01} and \cite{burlaga02} that during the interaction process the CMEs may merge and become a single magnetic structure. Using observations from STEREO for the 2007 January 24--25 events, \cite{lugaz08} and \cite{webb09} reported the overtaking of a slower CME by a faster CME such that the shock wave driven by the faster CME traveled through the slower preceding CME \citep{lugaz09}. The studies by \cite{lugaz08,lugaz09} and \cite{webb09} revealed that knowing the propagation direction of the potentially interacting CMEs is crucial in order to interpret the observations in a reliable way. Furthermore, enhanced radio emission \citep{gopalswamy01} and highly energetic solar particles have been reported in connection with a CME-CME interaction \citep[e.g.,][]{kahler94,kahler01,gopalswamy02}. Numerical simulations have shown, this is either due to the reconnection processes between the intrinsic magnetic fields of the CMEs, or the additional particle acceleration due to the associated shock wave propagating through the slower CME \citep[e.g.,][]{schmidt04,lugaz05}. All these results provide evidence that the intrinsic magnetic field of a CME, i.e.\ the assumed flux rope structure \citep[e.g.,][and references therein]{chen96,low01}, plays an important role with respect to its kinematical evolution.

As a CME propagates into interplanetary (IP) space it interacts with the ambient medium and transfers momentum and energy in form of MHD waves \citep[e.g.,][]{jacques77}. The interaction of the CME with the solar wind results in the adjustment between the speed of the CME and the solar wind flow. Assuming that the main force that governs the propagation behavior of a CME in IP space is the ``aerodynamic'' drag force \citep{cargill96,vrsnak01CME,vrsnak02gopal,cargill04}, we may attempt to simulate the kinematical profile of a CME by using the drag-based model (DBM) proposed by \cite{vrsnak07zic}, and advanced by Vr\v{s}nak et al.\ (2012; submitted to Solar Physics).

In the following study we focus on two CME events from 2010 August 1 and we investigate if the CMEs may have interacted with each
other as they propagated away from the Sun. We use STEREO/SECCHI observations to derive the three-dimensional properties of both CMEs (direction of motion, kinematical profile, width). The derived de-projected kinematical profile is then compared to the results from the DBM. From this we will show that (1) both CMEs head in similar directions as they evolve into IP space, making a CME-CME interaction highly likely; (2) a strong deceleration of the faster CME can be explained by interaction with the slower CME which can be represented as an ambient medium of higher density and flow speed compared to the unperturbed solar wind; (3) the interaction between the magnetic structures of the CMEs starts earlier than their leading edges merge in the heliospheric images.

\section{Data and Observations}\label{data-section}

For the CME-CME interaction study we analyze white light images from the Solar Terrestrial Relations Observatory (STEREO-A and STEREO-B) SECCHI instrument suite \citep{howard-stereo08}. SECCHI consists of two coronagraphs, COR1 and COR2, covering a plane-of-sky (POS) distance range up to $\sim$15~R$_\odot$, and the heliospheric imagers (HI), HI1 and HI2, for distances $>$15~R$_\odot$. This instrument combination allows us to track CME/ICME events in the inner heliosphere from $\sim$2~R$_\odot$ to beyond 1~AU. For context information on the low-coronal conditions close to the solar surface, we use data from the SECCHI Extreme Ultraviolet Imager \citep[EUVI;][]{wuelser04}.

Due to an observational gap, STEREO-B data are not available from $\sim$10:00 UT on 2010 August 1 until 04:00 UT on 2010 August 2. For that reason, techniques based on stereoscopic data are used close to the Sun but not for studying the entire propagation path up to 1~AU.

\begin{figure*}
\epsscale{0.8}
 \plotone{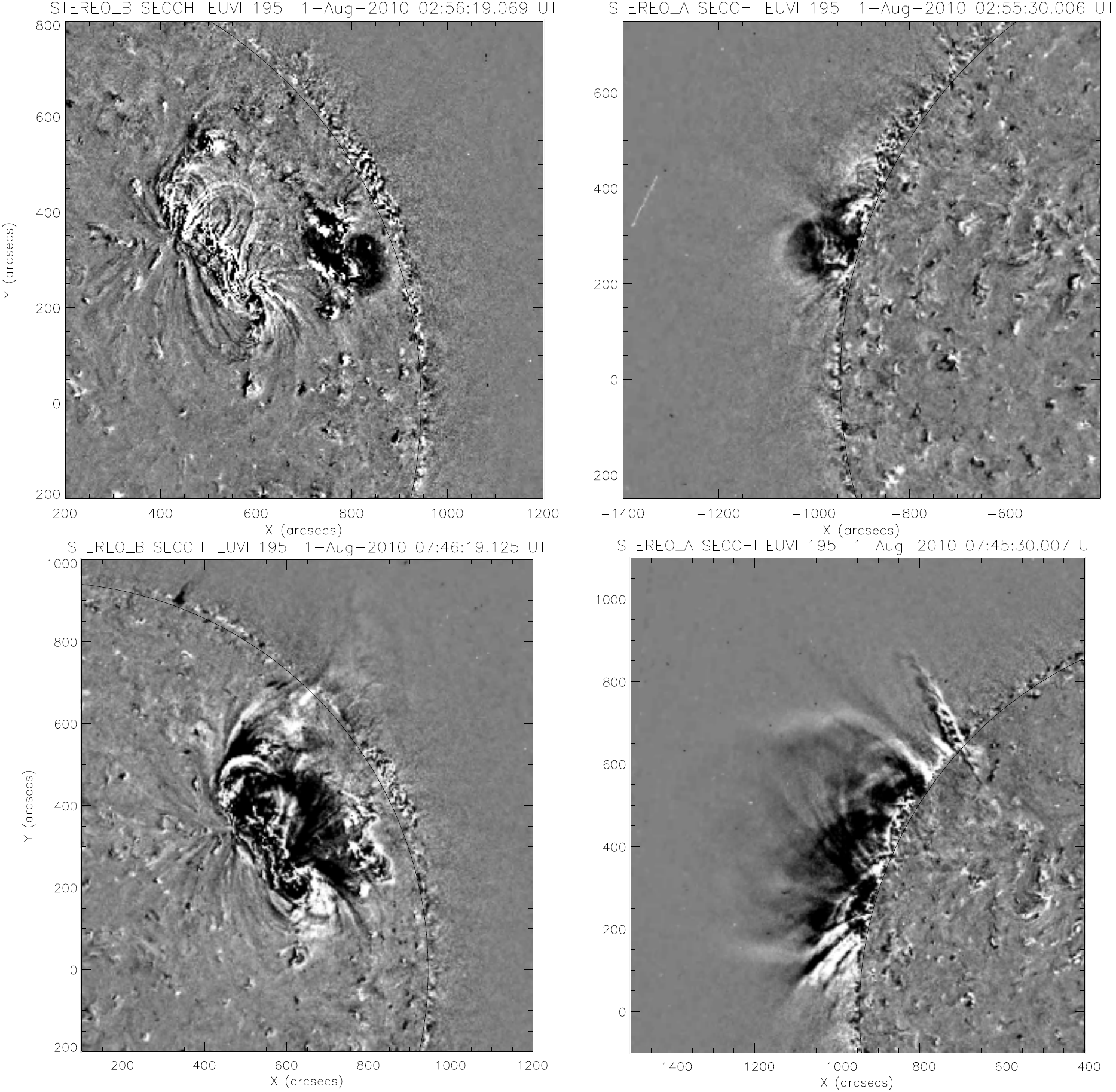}
 \caption{STEREO-A (right) and -B (left) EUVI 195~\AA~difference images showing the coronal dimming regions of CME1 (top) and CME2 (bottom). }
    \label{EUVI}
\end{figure*}

\begin{figure*}
\epsscale{.9}
 \plotone{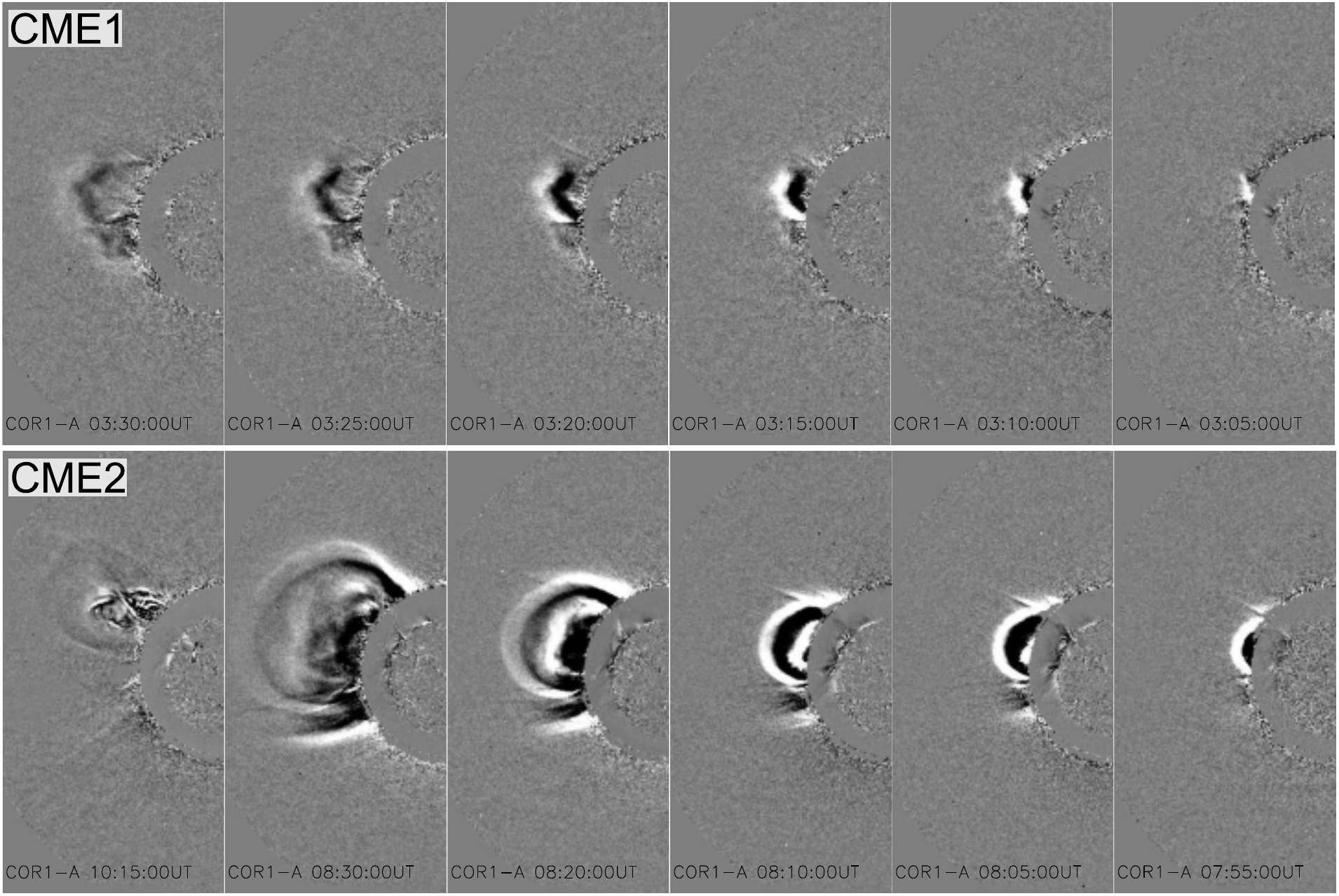}
 \caption{CME1 and CME2 in EUV and white light observations. {\it Top}: Sequence of EUVI/COR1 STEREO-A running difference images of CME1. {\it Bottom}: Sequence of EUVI/COR1 STEREO-A running difference images of CME2.}
    \label{SR-CME1+2}
\end{figure*}

The first CME (hereinafter CME1) was launched from the Sun on 2010 August 1 at $\sim$2:55~UT with a de-projected mean speed of $\sim$700~km~s$^{-1}$ in the COR1 and COR2 fields of view (FoVs). STEREO-A EUVI images (Figure~\ref{EUVI}, top right) show an off-limb dimming region, representing the low coronal signatures of CME1. From the solar surface signatures we also derive the location of source region (SR) for both CMEs as given in Table~\ref{correl} \citep[cf.][]{cremades04}. As can be seen from Figure~\ref{SR-CME1+2}, the CME feature can be seamlessly tracked as it propagates into the coronagraph FoV from which it can be further followed into the HI1 FoV. The CME was associated with a prominence eruption followed by a GOES B4.5 class flare in a small active region \citep[NOAA 11094; cf.][]{schrijver11} located at E14/N14 (if not stated otherwise all locations are heliographic coordinates given in degrees with respect to Earth).

The second CME (hereinafter CME2) was associated with a long duration flare classified as GOES C3.8 and was launched from AR 11092 located at about E35/N20. The flare started at $\sim$7:24~UT, lasted until 10:25~UT and had its maximum intensity at 08:56~UT \citep[for more details on the flare and associated filament eruption see][]{liu_r10,schrijver11}. CME2 was a fast event with a de-projected mean speed of $\sim$1200~km~s$^{-1}$ in the COR1 and COR2 FoVs. Clear low coronal signatures of CME2 are seen in STEREO-A EUVI at $\sim$7:45~UT (see bottom right panel of Figure~\ref{EUVI}). CME2 can be seamlessly tracked from EUVI to COR1 (Figure~\ref{SR-CME1+2}) where it shows a very distinct leading edge and is associated with a remote streamer deflection, probably related to a shock front \citep[cf.][]{liu09,gopalswamy09shock,ontiveros09}.

The in situ signature of the leading edge of CME2 was an interplanetary shock and its high-density sheath region recorded at 1~AU on 2010 August 3~17:05~UT \citep[cf.][]{harrison12} by Wind/SWE/MFI \citep{ogilvie95,lepping95}. The arrival time of the shock and proton bulk speed following its arrival are used to further constrain the kinematical profile of CME2 (see Section~\ref{R+IS}). The solar wind speed diagnosed in situ near Earth was around 400~km~s$^{-1}$ prior to the arrival of the shock associated with CME2 on 2010 August 3. The solar wind speed measured in the sheath region of CME2 increased to 600~km~s$^{-1}$ \citep[see Figure~14 in][]{harrison12}. A detailed study focusing on the in situ signatures of these events is given by M\"ostl et al.\ (2012, in preparation for ApJ). The readers are also directed to \cite{liu12} for connections between imaging observations and in situ signatures with
respect to the CME-CME interaction.

CME2 was associated with a type II radio burst showing distinct morphological changes at $\sim$09:50~UT on 2010 August 1. Emission at the fundamental frequency divided into two branches of different frequency drifts, implying the simultaneous existence of radio emission sites moving with different speeds. More details on the radio signatures associated with the 2010 August 1 events are given by Martinez Oliveros et al.\ (2012; in press for ApJ).

\section{Methods}\label{methods-section}

\subsection{Observations and reconstruction methods}\label{HM-method}

In the following section, we give details on different reconstruction methods used to obtain the three-dimensional characteristics of CME1 and CME2. Most important, we derive the direction of motion for both CMEs in order to give evidence of an actual interaction. For more details on different reconstruction techniques for stereoscopic data of CMEs in the early phase of the STEREO mission we refer the interested reader to \cite{mierla10}.

\subsubsection{Forward modeling}
The flux rope forward fitting model developed by \cite{thernisien06} and \cite{thernisien09} is a raytrace simulation method that computes synthetic total and polarized brightness images using the Thomson scattering formulae from an assumed electron density model. The appearance of a CME is approximated by a flux-rope like structure which is simulated by the graduated cylindrical shell (GCS) model. By fitting the density model to contemporaneous image pairs from STEREO-A and STEREO-B, which observe the CME from two different vantage points, we derive characteristic CME parameters like width and source region position, i.e.\ location of the CME apex projected back along a straight line normal to the solar surface \citep[][]{thernisien11}. For the reconstruction we chose those image pairs with the highest contrast and applied the forward modeling to imagery data from COR1 for CME1 and COR2 for CME2.

\subsubsection{Geometric triangulation}
The geometric triangulation technique developed by \cite{liu10a,liu10b} can convert elongation measurements to radial distance and propagation direction, taking advantage of stereoscopic imaging observations from STEREO-A and STEREO-B. The basis of this technique is that the propagation direction and distance yield a certain elongation angle corresponding to the viewpoint. The two viewpoints from STEREO then form a simple geometry with which the propagation direction and radial distance can be derived. The advantage of this technique is that it has no free parameters and does not assume a constant propagation direction and speed. This is of particular importance in the case of CME-CME interactions, as both the propagation direction and speed can be changed by the interactions. \cite{liu10a,liu10b} describe the mathematical formulas and detailed procedures for applying this technique. Here we apply the technique to CME1 and CME2 before their tentative collision. Because of a data gap in STEREO-B, the technique can be applied to COR2 and HI1 for CME1 and only to COR2 for CME2 \citep[cf.\ Figure~5 in][]{harrison12}.

\subsubsection{Fitting method for Harmonic Mean and Fixed Phi}
We also make use of two well-established fitting techniques, namely the so-called Fixed Phi (FP) fitting method developed by Sheeley and co-workers \citep{sheeley99,kahler07webb,rouillard08,sheeley08a} and the Harmonic Mean (HM) fitting method developed by \cite{lugaz10} to find constant propagation directions and speeds. The former technique effectively assumes that we are tracing a point-like source along a particular radial line from the Sun, the latter assumes a circular geometry for the CME. The fitting techniques are applied to elongation-time profiles derived from single spacecraft observations from both HI1 and HI2 for CME1 and CME2 under the assumption that the CME is moving with constant speed and in a constant direction. For a detailed application of this method to the data of 2010 August 1 we refer to \cite{harrison12}.

\subsubsection{Remote sensing and in situ data}\label{R+IS}
By combining observations from remote sensing instruments on a single spacecraft and in situ observations of CME signatures at 1~AU, we are able to estimate the direction of motion of a CME as well as its de-projected kinematics \citep{moestl09a,moestl10,rollett11}. The time-elongation profile of CME2 for the distance range from $\sim$15~R$_\odot$ to Earth is obtained from a elongation-time map made from STEREO-A HI1 and HI2 observations extracted along the ecliptic plane \citep[so called J-maps;][]{sheeley08a,davies09}. We convert the elongations into radial distances by assuming a constant direction of motion from Sun to Earth \citep{moestl09a} and applying the HM conversion method, i.e.\ assuming a circular geometry for the CME \citep[e.g.,][]{howard09,lugaz09}. The measured arrival time and speed of the shock and its following high-density sheath region at 1~AU are boundary conditions for the conversion method and constrain the possible range for the direction of motion of the CME under study. From this we obtain two directions, one that matches the arrival time and one that matches the arrival speed \citep[for more details on the method see][]{rollett11}. We note that the average of these two directions is taken as final result. Another boundary condition close to the Sun is represented by the kinematical profile of CME2 as derived from COR2 observations which slightly overlap the FoV of HI1. By combining distance-time and speed-time profiles up to $\sim$15~R$_\odot$ and at 1~AU with results from the conversion method which covers the distance range that lies in between, we derive the full kinematical profile of CME2 from Sun to Earth \citep[see also][]{temmer11}.

This method is applied solely to CME2 since the signature of CME1 can not be tracked to the distance of 1~AU in remote sensing images. Furthermore, only the shock driven by CME2 could be clearly identified at 1~AU from in situ Wind data. \cite{liu12} suggest that both CME1 and CME2 were observed in situ, however, this cannot be completely confirmed from the data.

\subsubsection{Total mass}
CMEs observed from different vantage points appear differently in their intensity distribution and thus apparent morphology, which is basically due to the line of sight integration of the white light emission from optically thin structures. The differences in the total intensity are due to the different incident angles of the Thomson scattering geometry through the CME plasma \citep[see][]{colaninno09}. Combining observations from both STEREO spacecraft enables us to estimate the direction of motion of the CME as well as its total (``true'') mass \citep{colaninno09}. The total mass is calculated for CME1 as well as CME2 by using stereoscopic image pairs from COR1 and COR2.

\subsubsection{Polarimetric localization}
Applying the polarimetric localization technique \citep{deKoning11}, the percent polarization observed by a single coronagraph is used to obtain a three-dimensional reconstruction of a whole CME. Polarimetric localization is based on the equations of \cite{billings66}, in which the elongation angle of a scattering point and the measured percent polarization can be related to the distance that the scattering point is from the plane-of-sky. Two possible solutions are derived from this method, one ahead of the spacecraft plane-of-sky, and one behind the plane-of-sky. The polarimetric localization solution from STEREO-A which is collocated with the one from STEREO-B gives the correct solution. From the whole CME reconstruction, the direction of motion as well as the kinematics of the CME in three-dimensional space is estimated. This method is applied independently to COR1 and COR2 images from STEREO-A and STEREO-B for both CME1 and CME2.

\subsection{Simulation of kinematics of CME2 using the DBM}\label{DBM-method}

The DBM (Vr\v{s}nak et al., 2012; submitted to Solar Physics) is based on the assumption that the kinematical profile of a CME in IP space is mainly controlled by drag force \citep{cargill96,vrsnak01CME,vrsnak02gopal,cargill04,vrsnak07zic}. The drag force can be expressed in its simplest form as $$F_{D}=\gamma(v-w)|v-w|~,$$ with $w$ the solar wind speed and $v$ the CME speed. The drag parameter $\gamma$ is defined as $$\gamma=C_{d} \frac{A_{\rm CME}\rho_{\rm sw}}{m_{\rm CME}}$$ where $C_{d}$ is the drag coefficient, $A_{\rm CME}$ the cross sectional area of the CME, $\rho_{\rm sw}$ the density of the solar wind, and $m_{\rm CME}$ the mass of the CME \citep[cf.][]{cargill04}. $C_{d}$ is a dimensionless number, typically of order of unity \citep[see][]{batchelor67}. Working on this assumption, we calculate the kinematical profile of CME2 and compare it with the kinematical profile derived from observations.

As input parameters, DBM requires the launch time $t_{0}$ and speed of the CME $v_{0}$ at a certain distance from the Sun $R_{0}$, the asymptotic solar wind speed $w$ at 1~AU as well as $\gamma$. From white-light observations, we derive $t_{0}$, $v_{0}$, and $R_{0}$, while $w$ is taken directly from Wind data. For $\gamma$ we obtain the parameters $A_{\rm CME}$ and $m_{\rm CME}$ from stereoscopic reconstruction methods, and $\rho_{\rm sw}$ is calculated using the empirical formula developed by \cite{leblanc98} depending only on the radial distance from the Sun. Assuming that a CME expands in a self-similar manner and that the density falls off as distance squared, $\gamma$ can be kept constant for the distance range from Sun to Earth. For a detailed description of the DBM we refer to Vr\v{s}nak et al.\ (2012; submitted to Solar Physics). The DBM is available for public usage through a web-interface under \url{http://oh.geof.unizg.hr/CADBM/cadbm.php}.

\section{Results}\label{result-section}

\begin{figure*}
\epsscale{1.}
 \plotone{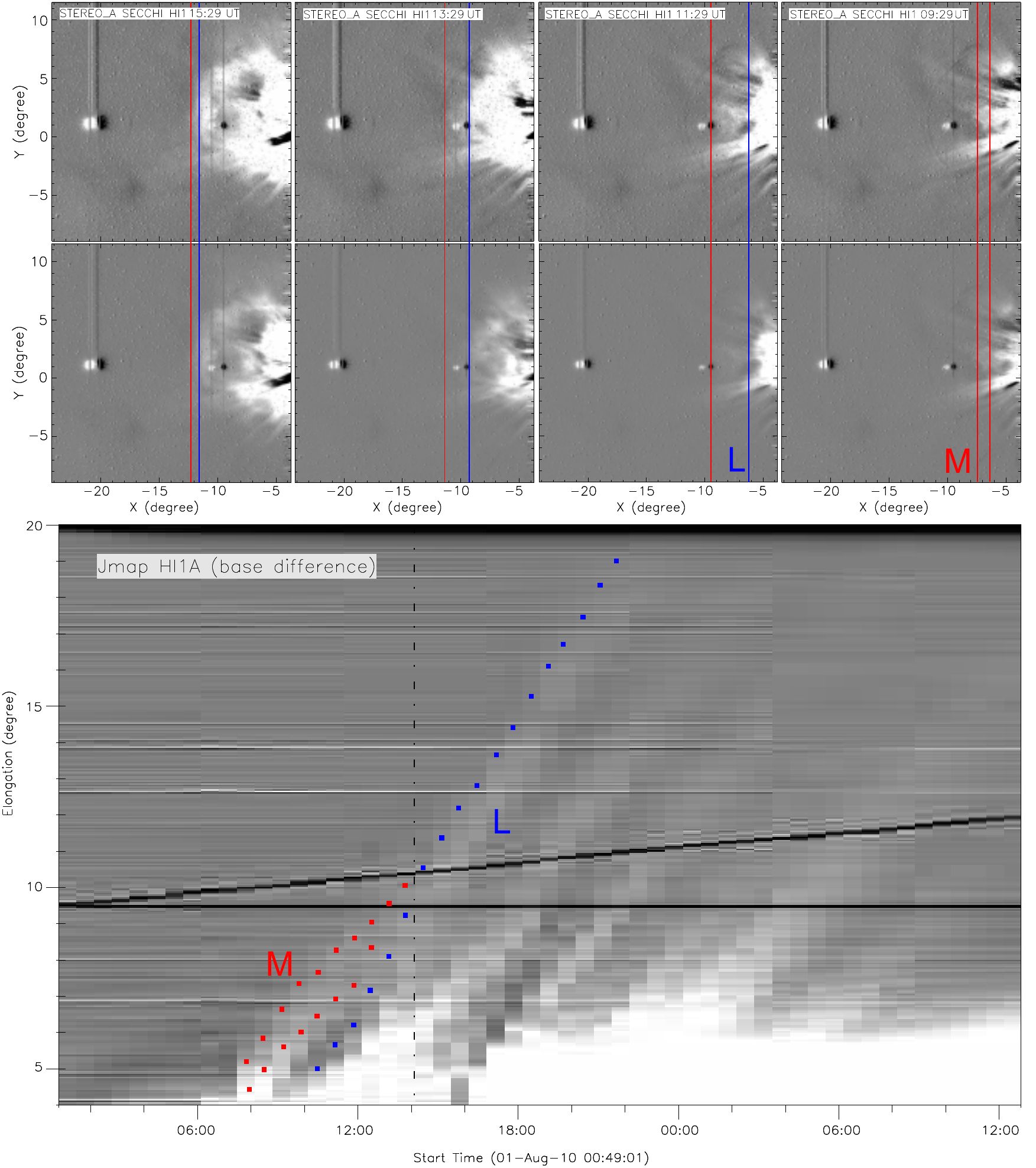}
 \caption{{\it Top}: Base difference images from HI1-A showing the evolution of CME1 and CME2 for the distance range $\sim$20--40~R$_\odot$. The frontal parts of CME1 (M) and CME2 (L) are indicated by red and blue vertical lines, respectively. Different contrast is used for top and bottom panels in order to better visualize CME1 which is fainter (less massive) than CME2. See also the accompanying movie. {\it Bottom}: J-map constructed from base difference images and overplotted tracks of CME1 (M; red squares) and CME2 (L; blue squares).}
    \label{ML}
\end{figure*}

Figure~\ref{ML} shows base difference images and J-maps for HI1-A data from which we observe that the signatures of the leading edge of CME1 and CME2 merged somewhere between 14:00 and 16:00~UT, corresponding to a distance of $\sim$38$\pm$5~R$_\odot$. After the interaction, no identifiable signatures of CME1 were observed, i.e.\ CME1 was ``lost'' in the white light signature of CME2.

Figure~\ref{forward-CME1} shows the simulated flux rope model (see Section~3.1.1) overlaid on white light COR1 (for CME1) and COR2 (for CME2) images. Applying this model, we derive the radial back-projection of the CME apex onto the solar surface which represents the launch site of the CME (CME1: E20/N09; CME2: E28/N20), the tilt angle relative to the solar equator (CME1: 39$\degree$; CME2: 47$\degree$), the face-on width (CME1: $\sim$68$\degree$; CME2: $\sim$100$\degree$), and the edge-on width (CME1: $\sim$23$\degree$; CME2: $\sim$52$\degree$). Typical uncertainties lie in the range of $\pm$10$\degree$ and depend on the identified boundaries of the CME.

The top panel of Figure~\ref{polariz} shows how the \cite{billings66} equations are used to obtain a reconstructed point within a CME relative to the spacecraft plane of sky (see Section~3.1.6). The bottom panels show two possible reconstructed CMEs for each event based on STEREO-A percent polarization measurements, one ahead of the spacecraft plane-of-sky, and one behind the plane-of-sky. The correct reconstruction, which can be deduced by utilizing in addition STEREO-B percent polarization measurements, is the colored reddish-orange one traveling eastward of the Sun-Earth line (i.e.,\ ahead of the spacecraft plane of sky). What can be seen immediately is that both CMEs have a similar trajectory, and that CME2 is significantly larger than CME1.

Table~\ref{correl} gives a summary of results for the direction of motion and speed of CME1 and CME2, respectively, derived from using the different methods described in Section~\ref{HM-method}. We find that both ejecta propagate in similar directions which makes an interaction between both CMEs within the FoV of HI1-A highly likely. Due to the unknown geometry of the CMEs we are not able to determine which method delivers the most reliable result. Taking the average of all methods used, we derive for CME1 the direction E16$\degree\pm$7$\degree$ which is close to its source region located at E14 (cf.\ Figure~\ref{EUVI}). For CME2 we derive the average longitude of E23$\degree\pm$15$\degree$. A large deviation from its associated source region (E35) is obtained for CME2 from the remote sensing and in situ (R+IS) method. This gives an average direction of E13 (E12 if we match the derivative of the converted elongation with the speed of the sheath region and E14 if we match the averaged arrival time of the sheath region at the Wind spacecraft). Inspecting the direction of motion for CME2 derived from other methods (forward modeling, polarimetric localization, HM) we obtain $\sim$E30--E35 which would be more consistent with the location of the source region.

Figure~\ref{convert} shows speed and radial distance profiles for the leading edge of CME2 that have been derived from elongation measurements from COR1/COR2/HI1/HI2 ecliptic observations from STEREO-A using directions E13 and E30. The speed profile is derived by performing numerical differentiation of the distance-time data applying three-point Lagrangian interpolation. The error bars in the speed profile represent uncertainties in the measurements of the leading edge of CME2 and are calculated as the standard deviation of the derivative from errors in the distance-time data. As can be seen, starting at a distance of $\sim$20~R$_\odot$ a clear deviation especially in the distance profile is obtained, revealing that E30 would not fit the boundary conditions at 1~AU. For that reason we chose to use the direction E13 for conversion and to compare with the results from the drag model. We note that this might give hints towards some longitudinal deflection of CME2, which is also found by \cite{liu12}. However, we can not rule out artifacts in the direction finding due to the geometrical assumptions underlying the different methods.

Inspection of the speed profile of CME2 as given in the top panel of Figure~\ref{convert} reveals that there were two distinctly different stages in the deceleration of CME2. The first one, characterized by a rapid deceleration of $a$~$\approx$$-$40~m~s$^{-2}$, lasted from $\sim$10~UT until $\sim$14~UT. After that, deceleration decreased to $a$~$\approx$$-$2~m~s$^{-2}$. In Figure~\ref{interact} we show the de-projected radial distance profile of CME1 \citep[results taken from the geometric triangulation method; see][]{liu12} extrapolated using a polynomial fit of second order until the time of interaction with CME2. Although the exact time depends on the direction of motion used for converting elongation into radial distance the interaction takes place between approximately 13:30~UT and 15:30~UT. From this we are able to attribute the first deceleration stage to the interaction of CME2 with CME1, whereas in the second stage the deceleration is due solely to the interaction of the merged structure with the ambient solar wind. The kinematical profile shows that the rapid deceleration stage actually begins a few hours \textit{before} the leading edges of CME1 and CME2 merged in the white light data ($\sim$10~UT), which can be attributed to a finite thickness of CME1.

\begin{table}[ht]
\centering \caption{Direction ($dir$) and speed ($v$) of CME1 and CME2 derived from SECCHI observations by applying different methods for different FoVs. The source region (SR) location is determined from low coronal signatures of each CME as observed in EUVI data (see Figures~\ref{EUVI}~and~\ref{SR-CME1+2}). We present results from combined remote sensing and in situ data (R+IS) from STEREO-A and Wind, respectively, the forward modeling (forw), calculating the ``true'' mass from stereoscopic STEREO-A and STEREO-B images (mass), the polarimetric localization technique (polar), geometric triangulation (triang) and harmonic mean and fixed phi fitting method (FP/HM). For a description of each method see Section~\ref{HM-method}. }
\begin{tabular}{l|cc|cc|c|c|c} \hline \hline
Method & \multicolumn{2}{c|}{$dir_{\rm CME1}$}  & \multicolumn{2}{c|}{$dir_{\rm CME2}$} & $v_{\rm CME1}$ & $v_{\rm CME2}$ & FoV \\ \hline \hline
SR      &  E14          & N14       &   E35         & N20   & ---   &  ---  & EUVI  \\
R+IS    &  ---          & ---       &  E13$\pm$10   & ---   &  ---  & 720$\pm$80  &  HI1+HI2+Wind \\
forw    &  E20$\pm$10   & N9$\pm$5  &  E28$\pm$5    & N20$\pm$5    & 650$\pm$150  &  1160$\pm$200 &  COR1+2  \\
mass    &  E5$\pm$5     & ---       &   E6$\pm$5    & ---   &  740$\pm$140  & 1250$\pm$100  & COR1+2  \\
polar   &  E19$\pm$8    & N1$\pm$2  &   E41$\pm$5   & N22$\pm$2  &  616$\pm$26 & 1264$\pm$66 & COR1+2 \\
triang  &  E21$\pm$9    & ---       &  E20$\pm$5    & ---   &  732$\pm$350  & 1138$\pm$550 &  COR2+HI1 (A+B) \\
FP/HM   &  ---          & ---       &  E2/E36       & ---   &  ---  & 764/960  &  HI1+HI2  \\

\end{tabular}
\label{correl}
\end{table}

From the results presented in Table~\ref{correl}, CME1 has a mean speed of $\sim$700~km~s$^{-1}$ within the FoV of COR2 and a total (``true'') mass of $\sim$6--8$\times$10$^{15}$~g; in contrast CME2 is a much faster event with a mean speed of $\sim$1200~km~s$^{-1}$ within the FoV of COR2 and has a total mass of $\sim$2--3$\times$10$^{16}$~g.

Figure~\ref{drag} shows the speed-distance, speed-time, and distance-time profiles of the leading edge of CME2 as derived from COR1/COR2/HI1/HI2 ecliptic observations from STEREO-A. The conversion to radial distance uses the result from the remote sensing and in situ method with E13 as ecliptic longitude. The speed profile and error bars are derived in the same way as given for Figure~\ref{convert}. The middle and bottom panels of Figure~\ref{drag} show the speed-time and distance-time curves of CME2 overlaid with the results from the DBM simulating two different scenarios of CME evolution.

For calculating the kinematical profile of CME2 from the DBM we use $C_{d}$=1 fixed over the Sun-Earth distance range, and constrain the other parameters by observations (see also Figure~\ref{scenario}). The asymptotic solar wind speed $w$ at 1~AU before the arrival of CME2 was measured by Wind as $\sim$400~km~s$^{-1}$ \citep[see][]{harrison12,liu12}. There is no indication from in situ data that CME2 crossed a high speed solar wind stream during its propagation from Sun to Earth which would result in a strong variation of the background solar wind speed $w$ \citep[see also][]{temmer11}. From the observational results of CME2 we derive $t_{0}$ to be 2010 August 1 10:19~UT and $v_{0}$=1400~km~s$^{-1}$ at $R_{0}$=15~R$_\odot$. We first calculate the kinematical profile DBM1 assuming that CME2 moves in unperturbed solar wind conditions from Sun to Earth. The parameter $\gamma$ is kept constant over the Sun-Earth distance range and chosen to be $\gamma$=0.25$\times10^{-7}$~km$^{-1}$ representing a CME which moves in an unperturbed environment of lower density and lower speed than the CME itself \citep[see also the parameter study by][]{cargill04}. As can be seen, the results from DBM1 based on such an assumption do not match the observational results. In another approach we calculate DBM2 using the same values for $t_{0}$, $v_{0}$, and $R_{0}$ but changing $w$ and $\gamma$ at a distance of 35~R$_\odot$ which is the distance at which the interaction of the two CMEs is likely to be finished. With this we simulate a scenario in which a CME moves up to 35~R$_\odot$ in an ambient flow that is of higher speed and density than the unperturbed environment. Beyond 35~R$_\odot$, CME1 and CME2 move as single structure in an ambient flow of speed and density comparable to an unperturbed solar wind, having the cross-section of CME2 and the combined mass of CME1 and CME2. In order to match the observational results, for radial distances $\leq$35~R$_\odot$ we used $\gamma_{1}$=$1.65\times10^{-7}$~km$^{-1}$ and $w_{1}$=600~km~s$^{-1}$, and for radial distances $>$35~R$_\odot$ we used $\gamma_{2}$=$0.2\times10^{-7}$~km$^{-1}$ and $w_{2}$=400~km~s$^{-1}$ (cf.\ Figure~\ref{scenario}).

In order to reproduce the CME-CME interaction in DBM2 we used values that give $\gamma_{1}/\gamma_{2}$=8.25. According to the CME-CME interaction scenario as described above we may express $\gamma_{1}=(A_{\rm CME2} \rho_{\rm CME1})/(m_{\rm CME2})$ and $\gamma_{2}=(A_{\rm CME2} \rho_{\rm sw})/(m_{\rm CME1}+m_{\rm CME2})$ from which we obtain $$\frac{\gamma_{1}}{\gamma_{2}}=\frac{\rho_{\rm CME1} (m_{\rm CME1}+m_{\rm CME2})}{\rho_{\rm sw} m_{\rm CME2}}~.$$

We now compare $\rho_{\rm CME1}$ and $\rho_{\rm sw}$ at a distance of 35~R$_\odot$. Assuming a cone-model \citep[e.g.,][]{michalek06}, we calculate the volume of CME1 using the derived width of $\sim$68$\degree$. Using the total mass of CME1 of $\sim$6--8$\times$10$^{15}$~g implies $\rho_{\rm CME1}$ $\sim$1.0$\times$10$^{-21}$~g~cm$^{-3}$. Applying the model by \cite{leblanc98} for the solar wind density we obtain $\rho_{\rm sw}$ $\sim$4.5$\times$10$^{-22}$~g~cm$^{-3}$ \citep[cf.][]{vrsnak10}. This gives us $\rho_{\rm CME1}$ $\approx$ 2~$\rho_{\rm sw}$. From total mass calculations we obtain  $m_{\rm CME2}$ $\approx$ 3~$m_{\rm CME1}$. Beyond $\sim$10~R$_\odot$ the mass is assumed to remain constant \citep{colaninno09} from which we suppose that the ratio between the masses is constant. We note that this is a simplistic assumption since \cite{lugaz05} have shown from simulations that the mass of a CME might increase up to a distance of 1~AU. From this we derive $\gamma_{1}/\gamma_{2}\sim$3, which shows that there is discrepancy to the model results. We may speculate that, besides the aerodynamic drag effect, magnetic forces (i.e.\ magnetic tension and magnetic pressure gradients raised due to the interaction) significantly contributed to the enhanced deceleration of CME2.

\section{Discussion and Conclusion}\label{conclusion-section}

Applying currently available models and reconstruction techniques for observations of the CME events from 2010 August 1 we are able to analyze a CME-CME interaction process quantitatively. Keeping in mind that the three-dimensional reconstruction techniques are only approximate and the kinematic results may vary greatly \citep[see e.g.][]{lugaz10b}, we use several (independent) methods. The results derived for the three-dimensional propagation directions are in agreement and provide evidence that a full interaction takes place between two consecutively launched CMEs. From observations we derive that the faster CME (CME2) is about three times more massive and $\sim$30\% larger in volume than the slower preceding CME (CME1). The white light signatures of the CMEs obtained from heliospheric imagery show that CME1 seems to be ``lost'' within the structure of CME2 since there are no features that can be attributed to CME1 after interaction \citep[see also][]{harrison12}. What we observe in white light as the leading edge of CME2 after merging with CME1 most probably includes the mass of CME1. Whether the magnetic features really merge, i.e.\ reconnect, or whether CME1 stays as magnetic entity that is compressed and pushed forward by CME2, is not possible to derive from remote sensing data. In situ data give more information on this aspect and are discussed in detail by \cite{liu12}. The results by these authors show that the multi-point in situ observations are consistent with compression of CME1 rather than its disintegration.

\begin{figure*}
\epsscale{1}
 \plotone{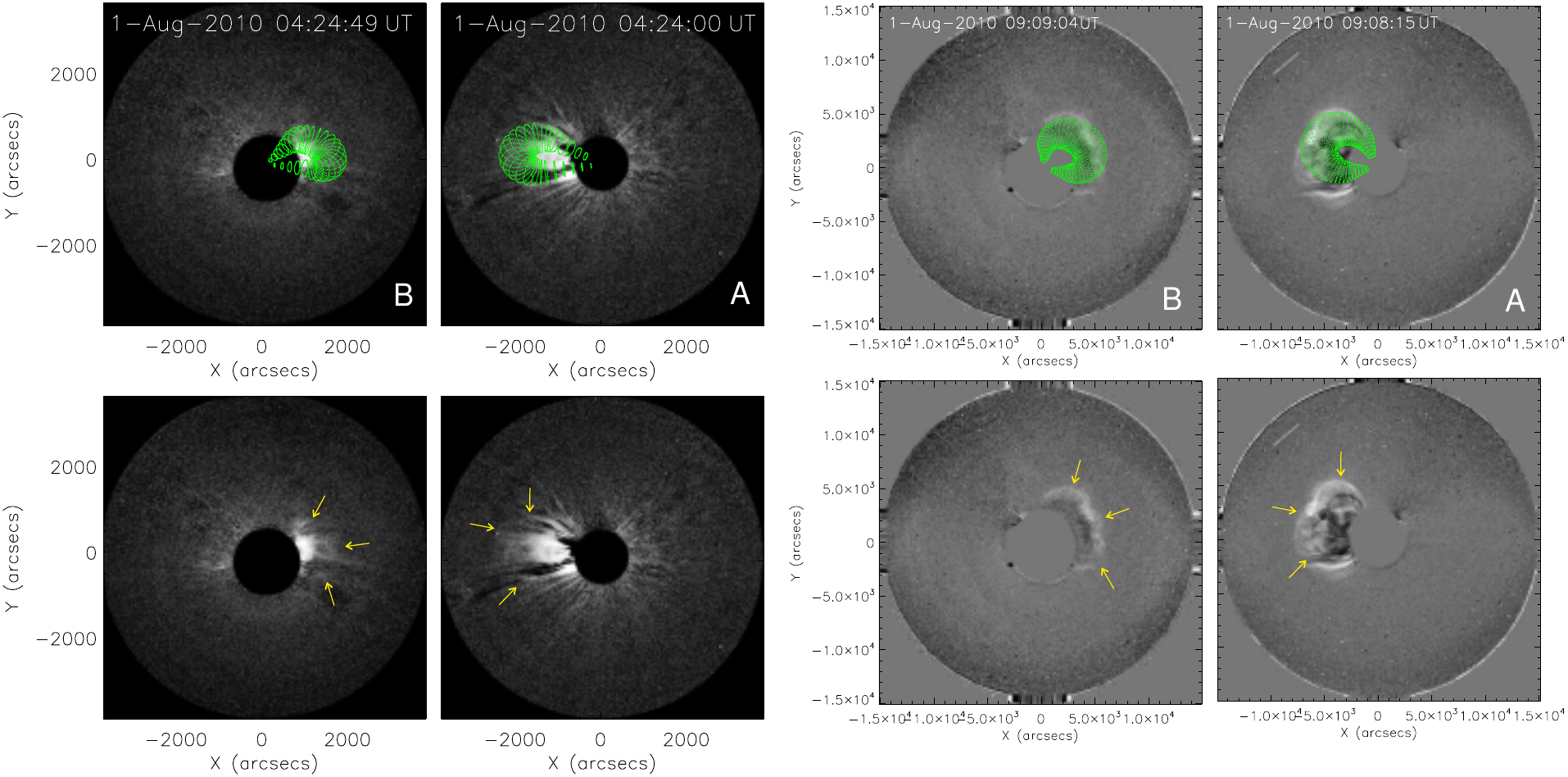}
 \caption{Results from forward modeling. {\it Left}: CME1 flux rope simulation fitted to white light data of COR1 aboard STEREO-A (A) and STEREO-B (B). The radial back-projection of the CME apex onto the solar surface gives a location of E20/N09. {\it Right}: CME2 flux rope simulation fitted to A and B white light data of COR2, yielding a surface location of E28/N20. The identified boundary of each CME is marked with yellow arrows in the bottom panels. }
    \label{forward-CME1}
\end{figure*}

\begin{figure*}
\epsscale{0.6}
 \plotone{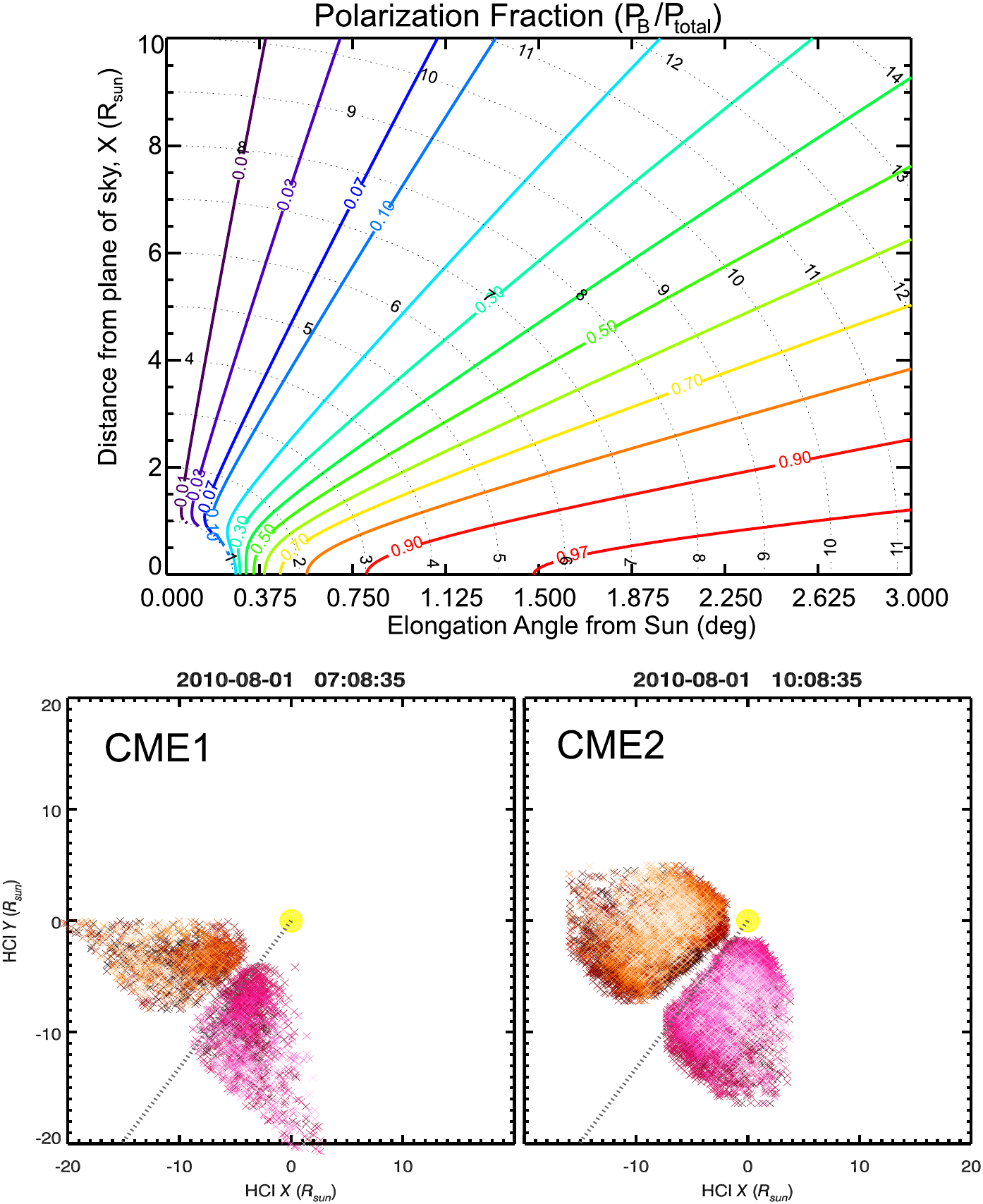}
 \caption{{\it Top}: Visualization of \cite{billings66} equations used to obtain a reconstructed point within a CME relative to the spacecraft plane-of-sky depending on the measured elongation and fractional polarization. {\it Bottom}: Each plot shows two possible reconstructed CMEs for CME1 (left) and CME2 (right) based on STEREO-A percent polarization measurements. The correct reconstruction is eastward of the Sun-Earth line, colored reddish-orange. The Sun-Earth line is shown as dashed line extending from the Sun marked as yellow circle.}
    \label{polariz}
\end{figure*}

\begin{figure*}
\epsscale{1.}
 \plotone{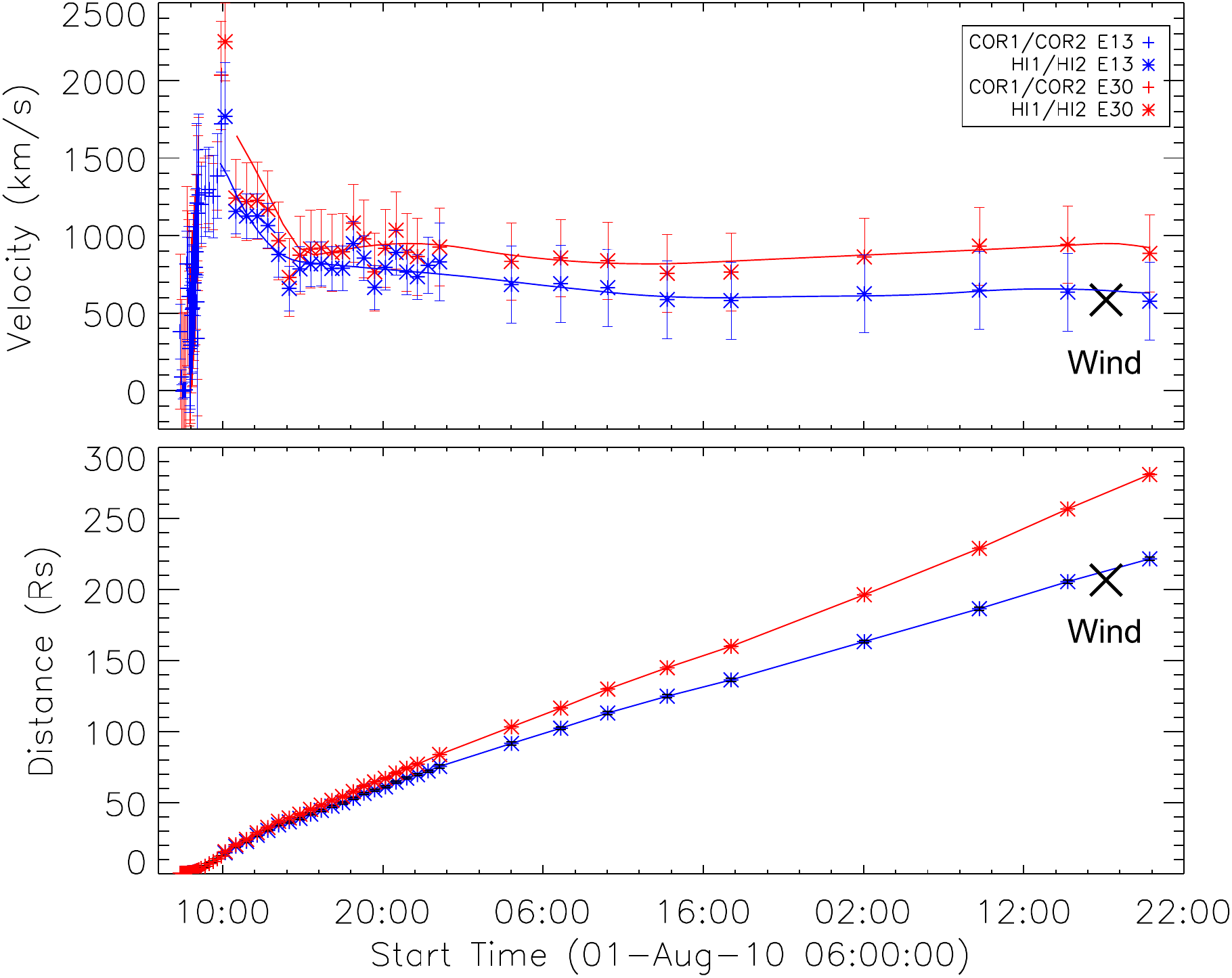}
 \caption{ {\it Top}: Velocity-time plot of CME2 using different directions for converting measured elongations into radial speed. Measured in situ speed of the sheath region from Wind is indicated with $\times$. {\it Bottom}: Distance-time plot of CME2 using different directions for converting measured elongations into radial distance. Measured arrival time of the shock at the Wind spacecraft is indicated with $\times$. The error bars are the standard deviation from the derivatives and represent the uncertainties in the measurements. }
    \label{convert}
\end{figure*}

\begin{figure*}
\epsscale{1.}
 \plotone{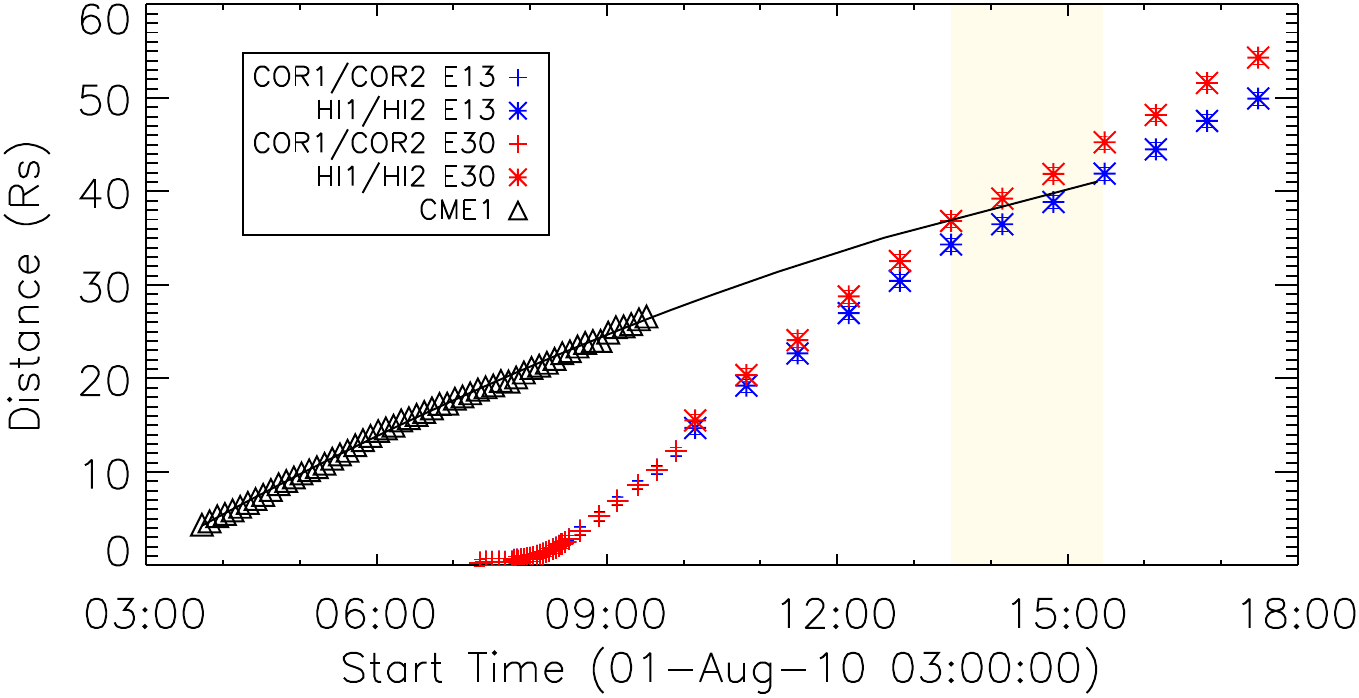}
 \caption{Velocity-time plot of CME2 using different directions for converting measured elongations into speed together with the kinematical profile of CME1 (geometric triangulation method) and its interaction time with CME2 (marked with a shaded box) using a polynomial fit of second order for extrapolation. }
    \label{interact}
\end{figure*}

\begin{figure*}
\epsscale{1.}
 \plotone{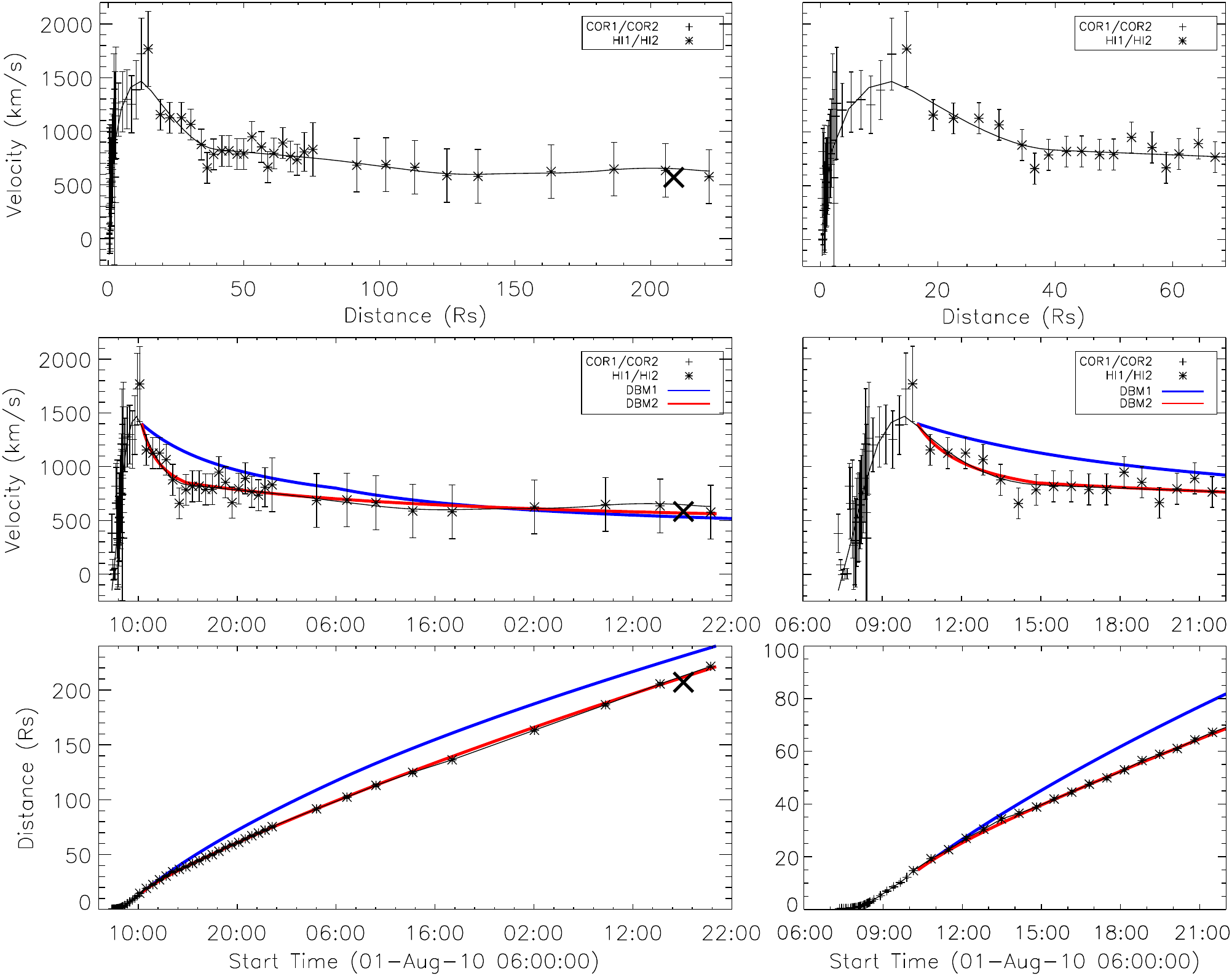}
 \caption{{\it Top}: Velocity-distance plot of CME2. {\it Middle}: Velocity-time plot of CME2 overlaid with the results from DBM1 (blue solid line; non-varying $\gamma$) and DBM2 (red solid line; varying $\gamma$ simulating the CME-CME interaction; see also Figure~\ref{scenario}). {\it Bottom}: Distance-time plot of CME2 overlaid with the results from DBM1 and DBM2. The black solid line is the spline fit to the observational data. The right hand panels are expanded portions of the left hand panels showing in detail the deceleration phase of CME2. Measured in situ parameters from Wind are indicated with $\times$ in the left hand panels. The error bars are the standard deviation from the derivatives and represent the uncertainties in the measurements.}
    \label{drag}
\end{figure*}

\begin{figure*}
\epsscale{1.}
 \plotone{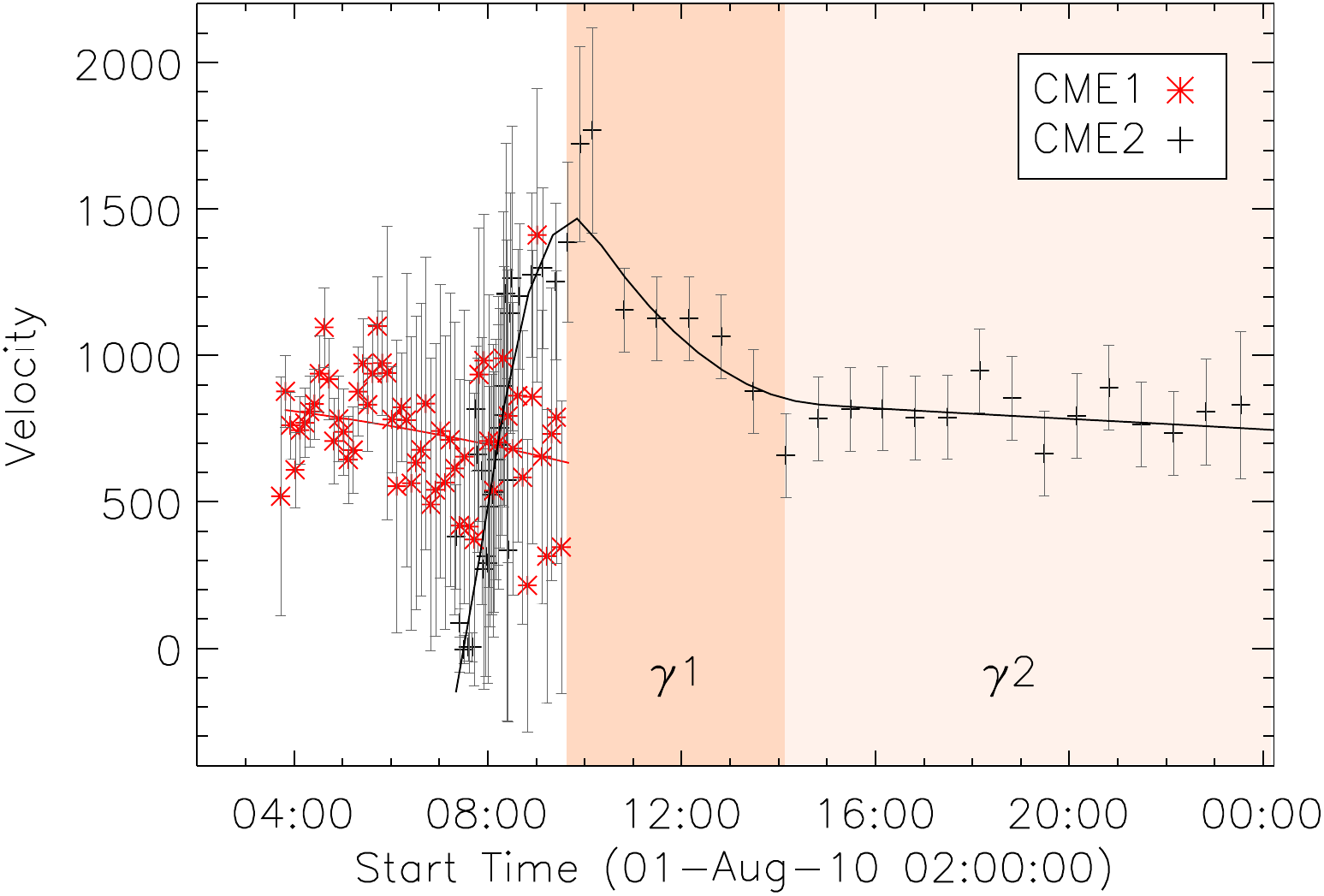}
 \caption{Velocity-time profile of CME1 (geometric triangulation method) and CME2. The shaded areas mark the range during which varying $\gamma$ values are applied. The error bars are the standard deviation from the derivatives and represent the uncertainties in the measurements. The red solid line represents a polynomial fit of second order to velocity data for CME1. The black solid line is the spline fit to the velocity data for CME2.}
    \label{scenario}
\end{figure*}

The CME-CME interaction is associated with transfer of momentum between the ejecta \citep{farrugia04} and this might be compared to simplified scenarios of one-dimensional elastic and inelastic collision \citep[cf.][]{lugaz09}. At the time of interaction at about 10~UT, the speed of CME1 is $\sim$600~km~s$^{-1}$ and CME2 $\sim$1400~km~s$^{-1}$, and the total mass of CME2 is derived to be $\sim$3 times higher than for CME1. Assuming a full interaction in the form of a perfectly inelastic collision, we would expect CME2 to move with a speed of $\sim$1200~km~s$^{-1}$. Assuming an elastic collision, CME1 would experience an acceleration, reaching a speed of $\sim$1800~km~s$^{-1}$. From observations, we derive a speed for the leading edge of the merged structure at about 15~UT of $\sim$800~km~s$^{-1}$ that makes both scenarios unlikely. As derived from the polarimetric localization method and the forward modeling (cf.\ Figures~\ref{forward-CME1} and \ref{polariz}), CME2 is significantly larger than CME1 from which it is not hard to imagine that CME1 will be ``lost'' within CME2. However, given the differences in size as well as mass, it is hard to imagine how the interaction between the two CMEs could result in substantial deceleration of CME2. Nevertheless, we must not forget that we are dealing with magnetic structures. As reported for the early kinematical evolution of CMEs, the interaction with strong overlying coronal magnetic fields may cause enhanced deceleration \citep[e.g.,][]{temmer08,temmer10}.

By applying the drag-based model (DBM) proposed by \cite{vrsnak07zic} and advanced by Vr\v{s}nak et al.\ (2012; submitted to Solar Physics), we are able to simulate the kinematical profile of CME2. The observational results can be reproduced by varying $\gamma$ values and ambient flow speeds in the drag formula at a radial distance from the Sun of $\sim$35~R$_\odot$ (distance at which the interaction process is likely to be finished). This can be interpreted in such a way that CME1 represents a magnetohydrodynamic obstacle for CME2 which leads to increased deceleration. In other words, CME2 propagates into a medium of denser plasma, higher flow speed, and stronger magnetic field than the unperturbed solar wind. After the interaction process the merged entity, i.e.\ CME2, carries the sum of masses of CME1 and CME2 and moves on as a single structure through an ambient flow of speed and density typical for quiet solar wind conditions.

Inspecting the kinematical profile of CME2 we derive that the strong deceleration starts a few hours before the merging of the white light leading edge signatures of the CMEs (10~UT versus $\sim$14~UT). This can be interpreted as an interaction between their magnetic structures, i.e.\ by a finite thickness of CME1. As shown at the end of Section~4, observations favor a scenario in which magnetic forces, induced during the CME-CME interaction, may play a substantial role in decelerating CME2. We speculate that the increase in magnetic tension and pressure when CME2 bends and compresses the magnetic field lines of CME1 increases the efficiency of drag. We would like to note that the interaction lasted for about $\sim$4--5 hours, which is probably too short for the total reconnection of the magnetic flux in CME1. This might support the results by \cite{liu12} who suggest that CME1 and CME2 most probably remain as two independent magnetic structures.

A metric type II burst emitted from two radio sources moving with different speeds is recorded at 09:50--10:15~UT, that is indicative of an interaction process \citep{gopalswamy01,gopalswamy02}. This supports the conclusion that the start of the interaction process is represented in the kinematical profile of CME2 by the strong deceleration observed around 10~UT. A detailed study for the 2010 August 1 events with respect to radio observations is given by Martinez Oliveros et al.\ (2012; in press for ApJ).

CME-CME interaction processes are still not well understood, mostly due to the lack of appropriate observational data to give information on the three-dimensional characteristics of CMEs. Interaction processes between ejecta are also important with respect to forecasting the arrival times of Earth directed CME events and their enhanced geoeffectiveness \cite[e.g.][]{burlaga87}. As was shown in the present study, the kinematical profile of a CME can be significantly changed due to the interaction with a preceding slower CME. Especially in times of enhanced solar activity CMEs can not be treated as isolated events. Analytical models in combination with unprecedented observations and three-dimensional reconstruction techniques are powerful tools which give us deeper insight and help us to better understand the evolution of CME events in IP space.

\acknowledgments We would like to thank Tim Howard for constructive discussions and an anonymous referee for helpful comments. M.T.\ greatly acknowledges the Austrian Science Fund (FWF): FWF V195-N16. The presented work has received funding from the European Union Seventh Framework Programme (FP7/2007-2013) under grant agreement n$\degree$ 218816 (SOTERIA) and n$\degree$ 263252 (COMESEP). B.B.\ was funded by the Austrian Space Applications Programme (ASAP-7 project 828271 3D-POC). This research was supported by a Marie Curie International Outgoing Fellowship within the 7th European Community Framework Programme. C.A.~de Koning was supported by NASA TR\&T grant NNX09AJ84G.

\newpage

\bibliographystyle{apj}

\end{document}